\documentclass[runningheads]{llncs}

\usepackage{makeidx}  % allows for indexgeneration
\usepackage{csquotes}
\usepackage{harveyballs}
\usepackage{tabularx}
\usepackage{caption}
\usepackage{subcaption}

\begin{document}
\mainmatter              % start of the contribution
%
% \title{Process Sustainability Patterns for Evaluating and Developing Sustainable BPM Technologies}
\title{Sustainability Analysis Patterns for Process Mining and Process Modelling Approaches}
\titlerunning{Sustainability Analysis Patterns}  % abbreviated title (for running head)
\author{Andreas Fritsch \orcidID{0000-0002-2124-4720}}
\authorrunning{Andreas Fritsch}   % abbreviated author list (for running head)
% \tocauthor{}
%
\institute{Karlsruhe Institute of Technology, Institute AIFB, Karlsruhe, Germany,\\
\email{andreas.fritsch@kit.edu}
}

\maketitle              % typeset the title of the contribution
% \index{Ekeland, Ivar} % entries for the author index
% \index{Temam, Roger}  % of the whole volume
% \index{Dean, Jeffrey}

% Todo: at most 150 words
\begin{abstract}     
Business Process Management (BPM) has the potential to help companies manage and reduce their activities’ negative social and environmental impacts. However, so far, only limited capabilities for analysing the sustainability impacts of processes have been integrated into established BPM methods and tools. One of the main challenges of existing Sustainable BPM approaches is the lack of a sound conception of sustainability impacts. This paper describes a set of sustainability analysis patterns that integrate BPM concepts with concepts from existing sustainability analysis methods to address this challenge. The patterns provide a framework to evaluate and develop process modelling and process mining approaches for discovering, analysing and improving the sustainability impacts of processes. It is shown how the patterns can be used to evaluate existing process modelling and process mining approaches.
\end{abstract}
\section{Introduction}
The demand for more sustainability is being put forward to companies from various sides, be it from politics, non-governmental organisations, customers or the company's workforce \cite{haldarConceptualUnderstandingSustainability2019}. %,ernstSMESReluctanceEmbrace2022}. 
The term sustainability or Sustainable Development stands for a long-term perspective to satisfy human needs today and in the future in light of escalating global challenges such as the climate crisis \cite{Brundtland1987}. 
%In achieving this goal, environmental protection and social justice are seen as central \cite{Brundtland1987}. 
From a company's point of view, contributing to sustainable development means considering the (negative) effects of its activities on humans (social justice) and the environment (environmental protection) \cite{Brundtland1987}. A consequence of this perspective is that the area of responsibility for a company expands \cite{heiskanenInstitutionalLogicLife2002}. It is not enough to consider only the direct impacts within the company's boundaries, such as wages to the company's workforce or pollutant emissions on its premises. Rather, the effects triggered by the company's activities along its value chain must also be considered \cite{isoISO1404020062006,unepGuidelinesSocialLife2020}. For example, 
%a company that operates a server also bears responsibility for the greenhouse gas (GHG) emissions caused by generating the energy needed to operate the server. % in a coal-fired power plant. 
% Similarly, 
a company that manufactures batteries bears responsibility for lithium mining and the associated contamination of %local communities' 
water resources \cite{nrdcLithiumMiningLeaving2022}. This idea is central to existing sustainability analysis methods, such as Life Cycle Assessment (LCA)  \cite{wbcsdGreenhouseGasProtocol2004,isoISO1404020062006,unepGuidelinesSocialLife2020}.
%, such as the GHG Protocol \cite{wbcsdGreenhouseGasProtocol2004} and Life Cycle Assessment (LCA) \cite{isoISO1404020062006,unepGuidelinesSocialLife2020}.

Business Process Management (BPM) concepts, methods, and tools have the potential to help companies become more sustainable.
BPM aims to analyse and improve the activities of companies \cite{weskeBusinessProcessManagement2019}. While traditionally improvements in terms of costs, lead times, and error rates are being sought \cite{dumasFundamentalsBusinessProcess2018}, researchers have begun to include sustainability aspects in their approaches \cite{couckuytGreenBPMBusinessoriented2019,fritschPathwaysGreenerPastures2022}. 
%These approaches can be subsumed under the term Sustainable BPM. 
% Adapting the definition of (conventional) BPM from , 
These Sustainable BPM approaches include concepts, methods and tools to support the modelling, analysis, improvement, implementation and management of business processes \cite{weskeBusinessProcessManagement2019,dumasFundamentalsBusinessProcess2018}, taking into account their environmental and social impacts. % TODO: could this be explicitly framed as a definition 
One of the challenges of existing Sustainable BPM approaches is the lack of a sound conception of how to measure the impacts of a business process \cite{couckuytGreenBPMBusinessoriented2019,fritschPathwaysGreenerPastures2022}. 

This paper provides a set of sustainability analysis patterns to address this challenge. The definition of the patterns is inspired by \enquote{workflow patterns} that have been used to evaluate process modelling and process mining approaches regarding their capabilities to represent control flow and data flow aspects of business processes \cite{russellWorkflowControlFlowPatterns2006,russellWorkflowDataPatterns2004}. Similarly, the sustainability analysis patterns describe different aspects of a process' environmental and social impacts. 
Note that the patterns are not intended as \emph{design} patterns, meaning they do not describe how an organisation can operate sustainably. 
Rather, they describe what aspects a business process \emph{model} should represent to provide a foundation for a sound, comparable and transparent sustainability analysis. 
This way, they guide (1) process analysts who aim to analyse the sustainability of business processes and (2) developers and researchers who integrate sustainability analysis capabilities in process modelling and process mining approaches. 
The patterns are derived by mapping a meta-model of BPM concepts with a meta-model of concepts from existing sustainability analysis methods (i.e. LCA). They are evaluated by applying them to compare the capabilities of existing process modelling and mining approaches and derive improvement possibilities for future developments. 
%The evaluation shows how the patterns can inform a critical appraisal of a Sustainable BPM approach's capabilities and identify potential for future developments.
%
% in Sustainable BPM.

% The paper is structured as follows: 
In Section~\ref{sec:background}, we provide background on BPM and LCA as well as an overview of related pattern proposals. Section~\ref{sec:mapping} provides a synopsis of BPM and LCA meta-models. Section~\ref{sec:patterns} presents the developed sustainability analysis patterns, and in Section~\ref{sec:evaluation}, the patterns are applied to a review of existing process modelling and mining approaches. Finally, we conclude in Section~\ref{sec:conclusion} and provide an overview of future work.

\section{Background and Related Work}
\label{sec:background}

\subsection{Business Process Management (BPM) and Process Mining}
\label{subsec:bpm}

BPM is concerned with improving 
%the performance of a company \cite{weskeBusinessProcessManagement2019,dumasFundamentalsBusinessProcess2018}.
business processes 
%(the activities carried out in a company \cite{oberweisModellierungUndAusfuhrung1996}), 
and thus the performance of a company \cite{weskeBusinessProcessManagement2019,dumasFundamentalsBusinessProcess2018}.
% To overview the multitude of BPM concepts, tools, and methods, 
A distinction can be made between management and technical BPM approaches \cite{couckuytGreenBPMBusinessoriented2019}. Management approaches address issues such as organisational structure, values, roles, and responsibilities within a company \cite{vanlooyConceptualFrameworkClassification2014}. Technical approaches are concerned with the modelling, analysis, improvement and implementation of business processes \cite{dumasFundamentalsBusinessProcess2018,weskeBusinessProcessManagement2019}.
% can be categorised using the business process life cycle concept \cite{couckuytGreenBPMBusinessoriented2019, vanlooyConceptualFrameworkClassification2014,dumasFundamentalsBusinessProcess2018}. 
% Literature has several definitions of the concept of the business process life cycle (e.g. \cite{dumasFundamentalsBusinessProcess2018, weskeBusinessProcessManagement2019}).
% The definitions vary in detail, but they all share the idea that business processes go through cycles in which they are continuously improved. 
% Following the definition given in \cite{dumasFundamentalsBusinessProcess2018} in a slightly adapted and simplified form, the business process life cycle consists of four phases: modelling, analysis, improvement, and implementation. In the first phase, modelling, a business process model is created. Based on the model, investigations can be carried out in the analysis phase, which in turn lead to improvements. The improved process model can be implemented, executed, and monitored in a supporting information system. Various (technical) approaches to BPM support one or more of the described phases. For example, modelling languages such as BPMN, EPC or Petri Nets support the modelling of a business process 
% \cite{omgBusinessProcessModel2013, scheerARISVomGeschaftsprozess2013, koschmiderPetriNetbasedView2018}. 
% \cite{koschmiderPetriNetbasedView2018}. 
% Modelling methods such as Horus \cite{schonthalerBusinessProcessesBusiness2012} or ARIS \cite{scheerARISVomGeschaftsprozess2013} also provide support for analysis, improvement, and implementation. 
Process mining is a branch of BPM \cite{vanderaalstProcessMining2016}. The basic idea of process mining is to automatically create business process models from event log data (e.g. events logged by IT systems) \cite{vanderaalstProcessMining2016}. %Various methods based on process mining (process discovery, conformance checking, …) support different phases of the business process life cycle \cite{vanderaalstProcessMining2016}. %In the following, we subsume process mining related approaches as well as general BPM concepts, methods and tools under the term BPM technologies.

\subsection{Sustainability Analysis and Life Cycle Assessment (LCA)}
\label{subsec:lca}

In the so-called Brundtland report, the term Sustainable Development was coined as a guiding principle for global change \cite{Brundtland1987}. %\cite{garrigaCorporateSocialResponsibility2004}. %Accordingly, the world should develop sustainably. 
The goal is to satisfy the needs of current generations without endangering the ability of future generations to satisfy their needs \cite{Brundtland1987}. The joint consideration of social, environmental, and economic factors is an essential aspect of the sustainability concept described in the Brundtland report 
%\cite{garrigaCorporateSocialResponsibility2004}, 
\cite{purvisThreePillarsSustainability2019}.

% Analysing a company's activities' social and environmental impacts represents a challenge. 
% A variety of different impacts must be included in a comprehensive sustainability analysis. 
It is important to consider that an activity's sustainability impacts may occur outside the company's boundaries. An important concept in this context is life cycle thinking \cite{heiskanenInstitutionalLogicLife2002}. It means that a systemic perspective is adopted when analysing the sustainability of a product or a company. This includes the interactions between the product or company under consideration, the environment, and other stakeholders along its value chain (also called life cycle) \cite{isoISO1404020062006}. The goal is to improve the entire system \cite{isoISO1404020062006}. %Life cycle thinking is seen as essential for investigating the sustainability of companies  \cite{elkingtonSustainableCorporationWinwinwin1994}. 
Life Cycle Assessment (LCA) is a method for conducting such an investigation \cite{isoISO1404020062006}.

% The term life cycle must be distinguished from the business process life cycle described in the Subsection~\ref{subsec:bpm}. In the case of LCA, a (product) life cycle primarily refers to the phases in a value chain. They range from the extraction of the raw materials needed for the product via the creation of the product and its use to the disposal or recycling of the product \cite{isoISO1404020062006}. In the case of a company life cycle, this consideration is extended to the inputs and outputs of an entire company \cite{isoISOTS140722014}. In the following, the term life cycle refers to the LCA concept. When referring to the BPM concept, the term business process life cycle is used.

The LCA method was initially developed to analyse the environmental impacts of products, e.g., contribution to climate change, acidification, ecotoxicity and waste \cite{klopfferLifeCycleAssessment2014,isoISO1404020062006}. 
% It aims to comprehensively assess different associated impacts. 
The method was adapted and extended for analysing the environmental impacts of an entire company \cite{isoISOTS140722014}, as well as social impacts \cite{unepGuidelinesSocialLife2020}. 
The corresponding standards and guidelines define that an LCA study consists of four phases \cite{isoISO1404020062006,isoISOTS140722014,unepGuidelinesSocialLife2020}. 
The first phase is the \emph{definition of goal and scope}. Here, the system boundaries, as well as basic requirements and assumptions, are defined. Second, \emph{inventory analysis} is carried out, in which data about the system is collected, and calculations are made to quantify relevant inputs and outputs. The third phase is the \emph{impact assessment}, where the significance of sustainability impacts is evaluated. And finally, as the last step, the \emph{interpretation} of the results, where improvement potential is identified. The GHG Protocol \cite{wbcsdGreenhouseGasProtocol2004} guidelines can be considered as a restricted variant of LCA since it only considers one impact category (climate change) \cite{unepGuidanceOrganizationalLife2015}. It largely follows the structure outlined by the LCA standards and provides definitions for different assessment scopes regarding GHG emissions \cite{wbcsdGreenhouseGasProtocol2004}: \emph{Scope 1} refers to direct GHG emissions within the premises of a company, \emph{Scope 2} refers to indirect GHG emissions that stem from the generation of energy consumed by a company, and \emph{Scope 3} addresses further indirect GHG emissions in the value chain.

\subsection{Related Work}

In Sustainable BPM literature, several proposals for sustainable or green business process patterns can be found \cite{Larsch2017,Lubbecke2016a,Nowak2011c}.
% ,schoormannElevatingSocialSustainability2019
These proposals for patterns describe, in a structured way, how the sustainability of a business process can be improved (e.g. using resources with reduced environmental impact \cite{Nowak2011c}). In each case, it is implicitly assumed that the impacts of a business process \emph{have already been measured}. The patterns described in this paper focus on the more fundamental question, \emph{how} the sustainability impacts of a business process should be measured.
%\section{Related Work}
%
%sbpm in general
%modeling approaches
%process mining approaches

\section{Mapping of BPM and LCA Meta-Models}
\label{sec:mapping}

\subsection{Meta-Models for BPM and LCA}

This section maps basic concepts from the BPM and LCA disciplines using meta-models. 
This way, a synopsis of concepts from both disciplines is constructed that allows for adapting insights from LCA for Sustainable BPM. 
From this synopsis, patterns that describe how the sustainability impacts of a process can be modelled are derived in the next section. 

\begin{figure}[htbp]
\centerline{\includegraphics[width=\linewidth]{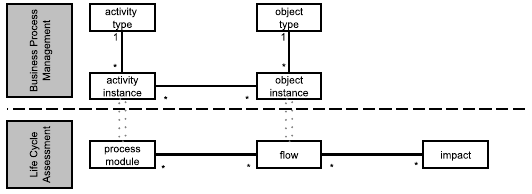}}
\caption{Mapping of BPM and LCA meta-models.}
\label{fig:datamodel}
\end{figure}

Fig.~\ref{fig:datamodel} shows a mapping between the meta-models of BPM and LCA. 
% The figure is intended as a basis for discussing basic concepts from both disciplines. It is not meant as a complete mapping 
From left to right, it shows similar concepts in the respective disciplines. As a basis for describing BPM concepts, we refer to the OCEL 2.0 standard \cite{bertiOCELObjectCentricEvent2023}. The choice is motivated by the concise form of the given meta-model and its interoperability between BPM technologies. Similarly, as a basis for describing LCA concepts, we refer to a description of the ILCD standard \cite{wolfInternationalReferenceLife2011}, which is used to exchange data between LCA tools. Note that in both cases, some of the terminology has been adapted for consistency and clarity within this paper. For example, the OCEL standard defines a concept \emph{event} as an execution of an activity. In Fig.~\ref{fig:datamodel}, this concept is called an \emph{activity instance}. In the following description of Fig.~\ref{fig:datamodel}, we additionally refer to basic sources of BPM \cite{weskeBusinessProcessManagement2019,oberweisModellierungUndAusfuhrung1996} and LCA \cite{isoISO1404020062006,klopfferLifeCycleAssessment2014} to explain the concepts and conclusions.

In BPM, the notion of a process refers to a business process, i.e. a set of \emph{activities} that are performed by a company to realise a business goal \cite{weskeBusinessProcessManagement2019}. Processes are modelled and analysed to improve a company's performance or to support the development of information systems \cite{weskeBusinessProcessManagement2019}. Activities may be associated with \emph{objects}: objects can be inputs or outputs of an activity, they may be altered by an activity, or they may be resources (machines, humans) that are needed to execute an activity \cite{bertiOCELObjectCentricEvent2023,weskeBusinessProcessManagement2019}. It is common to differentiate between \emph{activity and object types and instances} \cite{weskeBusinessProcessManagement2019,bertiOCELObjectCentricEvent2023}. Typically, a business process is modelled at the type level, where the model describes a general frame for executing the process \cite{oberweisModellierungUndAusfuhrung1996}. Activity instances refer to executions of an activity at a specific time \cite{bertiOCELObjectCentricEvent2023}. BPM traditionally focuses on causal and temporal relationships between activities (control flow) \cite{weskeBusinessProcessManagement2019,dumasFundamentalsBusinessProcess2018}. 
% The consideration of relationships between activities and objects (data flow) is a secondary concern for widely used process modelling languages such as BPMN or EPC \cite{weskeBusinessProcessManagement2019}. 
However, several approaches aim to integrate the data flow perspective more tightly  \cite{steinauDALECFrameworkSystematic2019}. 
Similarly, the early focus of process mining approaches was on the relationships between activities, with a recent uptake of interest in analysing the relationships between activities and objects  \cite{bertiOCELObjectCentricEvent2023}.

In LCA, process models are created to analyse the sustainability impacts of a product, service or company \cite{isoISO1404020062006,isoISOTS140722014,unepGuidelinesSocialLife2020}. Different to how processes are typically modelled in BPM, these process models represent the phases of a value chain (also called life cycle) and are built from \emph{process modules} \cite{klopfferLifeCycleAssessment2014,wolfInternationalReferenceLife2011}. A process module may represent a phenomenon similar to an activity in BPM (e.g. manufacturing of a workpiece), but it may also represent a whole industry (e.g. energy generation in Germany) \cite{klopfferLifeCycleAssessment2014}. Process modules are associated with %(sustainability-relevant) 
\emph{flows} \cite{wolfInternationalReferenceLife2011}. Similar to the object concept in BPM, flows may be products, information or data, but the focus is on material and energy flows  \cite{klopfferLifeCycleAssessment2014,isoISO1404020062006}. When modelling processes, LCA doesn’t distinguish between types and instances. Instead, a process model represents the quantified flows along a value chain relative to a \enquote{functional unit} (e.g. the flows required to provide a certain amount of mineral water) \cite{isoISO1404020062006}. 
The identified flows are then associated with different \emph{impacts} \cite{isoISO1404020062006,wolfInternationalReferenceLife2011}. 
The resulting impact indicators quantify the extent of a certain flow's impact on an environmental or social problem area \cite{isoISO1404020062006}.
For example, emissions of \emph{CO2} and other GHG emissions (\enquote{flows}) contribute to the problem area \emph{climate change} (an \enquote{impact}). 
This impact is typically measured in CO2 equivalents (CO2e) \cite{klopfferLifeCycleAssessment2014}. 
Other environmental impacts include ozone depletion, land use, ecotoxicity and acidification. 
Examples of impacts addressed in social LCA are excessive working hours, work accidents or discrimination \cite{unepMethodologicalSheetsSubcategories2021}. 
Note that any list of considered impacts depends on scientific progress and societal awareness \cite{klopfferLifeCycleAssessment2014}. For an overview of environmental impact lists, see \cite{klopfferLifeCycleAssessment2014}. For a list of impacts considered in social LCA, see \cite{unepMethodologicalSheetsSubcategories2021}.
% The methodological sheets for social LCA provide a comprehensive list of social impacts based on international norms.
% In every case, flows are classified and characterised to calculate quantified impact measurements \cite{klopfferLifeCycleAssessment2014,isoISO1404020062006}. 

\subsection{Specific Characteristics of Process Modelling in LCA}

Several notable aspects of process modelling in LCA are relevant when considering how to model sustainability aspects in BPM.
% In particular, the \emph{goal and scope definition} phase (see Subsection~\ref{subsec:lca}) significantly influences the resulting process models.
In particular, the relevant flows depend on the the assessed impacts.
For example, the assessment of climate change requires the identification of GHG emissions, while the assessment of ozone depletion requires the identification of gas emissions that damage the ozone layer \cite{klopfferLifeCycleAssessment2014}.

The scope definition is another important factor in the identification of relevant flows. For example, a \enquote{gate-to-gate} analysis only considers flows within a company \cite{isoISO1404020062006}. 
A \enquote{cradle-to-gate} analysis extends the scope of considered flows to include the upstream value chain of the company (the company's suppliers and their suppliers etc.) \cite{isoISO1404020062006}. 
Scopes 1 to 3, as defined in the GHG Protocol, are another example of a scope definition specific to climate change impacts \cite{wbcsdGreenhouseGasProtocol2004}. 

A further important aspect is the allocation of impacts. When dealing with process modules that have multiple inputs or outputs (e.g. mining produces diamonds and rubble), it needs to be decided how the associated impacts are distributed \cite{klopfferLifeCycleAssessment2014}. This is a central problem in LCA and conventions exist how to handle these situations (e.g. allocation by weight or economic value) \cite{isoISO1404020062006,klopfferLifeCycleAssessment2014}. In any case, allocation decisions can have significant consequences in the resulting assessment, so they must be carefully considered and documented \cite{isoISO1404020062006,klopfferLifeCycleAssessment2014}.

\subsection{Conclusion: Sustainability Impacts of a Business Process}

In both disciplines, BPM and LCA, processes are modelled and analysed. The observed similarities 
%(\emph{activities} are related to \emph{process modules}, \emph{flows} are related to \emph{objects}) 
allow for using LCA as a reference to reason about how to measure sustainability impacts in BPM \cite{couckuytGreenBPMBusinessoriented2019,fritschPathwaysGreenerPastures2022,gravesReThinkYourProcesses2023}. 
%Researchers have noted that Sustainable BPM should learn from sustainability analysis methods such as LCA \cite{couckuytGreenBPMBusinessoriented2019}, \cite{fritschPathwaysGreenerPastures2022}, \cite{gravesReThinkYourProcesses2023}. 
From the mapping conducted above, there are several lessons to be learned. First, Sustainable BPM requires some kind of object or data flow perspective to capture impacts (see also \cite{gravesReThinkYourProcesses2023}). 
% This case was also made in \cite{gravesReThinkYourProcesses2023} specifically for process mining. 
% If (relevant) objects are identified, they can be associated with impacts, similar to LCA flows.

Second, Sustainable BPM needs a conception of the scope of an analysis. Speaking in LCA terms, a gate-to-gate analysis may yield significantly different results than a cradle-to-gate analysis. Taking, for example, a business process supported by IT, it needs to be clarified if an analysis considers Scope 2 (electricity would be the main relevant input) or Scope 3 (the IT infrastructure and its impact along the value chain would have to be considered as well). 
%Note that a Scope 3 
%(or, generally speaking, cradle-to-gate) 
%analysis does not require explicitly modelling the whole value chain. The corresponding impacts may be quantified in aggregated indicators \cite{klopfferLifeCycleAssessment2014,wbcsdGreenhouseGasProtocol2004}.

Third, allocation is an important challenge to be addressed by Sustainable BPM. In the BPM modelling perspective, multi-input and multi-output processes are commonplace (e.g., a resource is used by multiple activity instances, an activity may have multiple object outputs). Thus, mechanisms for transparently and consistently allocating impacts are required.

\section{Sustainability Analysis Patterns}
\label{sec:patterns}

% Varying complexity and expressiveness

Based on the results of Section~\ref{sec:mapping}, the following sustainability analysis patterns for business processes are defined. They describe different sustainability-relevant aspects of business processes and adapt concepts of LCA for BPM. 
Their implementation in approaches for Sustainable BPM thus enables the analysis of process sustainability from a life cycle perspective. 
For the four patterns, 
%\emph{(AP~1) Assignment of sustainability-relevant inputs and outputs}, \emph{(AP~2) Assignment of sustainability impacts}, \emph{(AP~3) Scoping of sustainability impacts}, and \emph{(AP~4) Allocation of sustainability impacts}, 
a description, a motivation and variants (if applicable) are provided. 
% The formulation and description of the sustainability patterns are based on patterns that have been formulated in the literature for various business process phenomena (for example, control flows \cite{russellWorkflowControlFlowPatterns2006}, and data flows \cite{russellWorkflowDataPatterns2004}. 
% They are formulated in a generic way, regardless of specific process modelling languages or tools. 
In the following, the term \emph{process component} refers to any of the BPM concepts shown in Fig.~\ref{fig:datamodel}.
% Similar to \cite{russellWorkflowControlFlowPatterns2006}, \cite{russellWorkflowDataPatterns2004} we use the term process component meaning any of the BPM concepts in figure XXX. In this way, the patterns are generally formulated so that they can be applied to different modeling approaches of business process management, regardless of specific process modeling languages or tools

\begin{figure}
     \centering
     \begin{subfigure}[b]{0.22\textwidth}
         \centering
         \includegraphics[width=\linewidth]{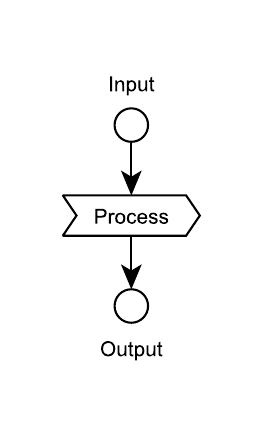}
         \caption{AP~1}
         \label{fig:pattern-1}
     \end{subfigure}
      \hfill
     \begin{subfigure}[b]{0.22\textwidth}
         \centering
         \includegraphics[width=\linewidth]{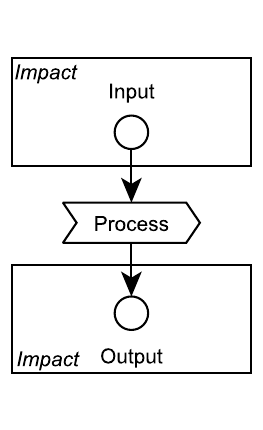}
         \caption{AP~2}
         \label{fig:pattern-2}
     \end{subfigure}
      \hfill
     % \vspace{.2cm}
     \begin{subfigure}[b]{0.22\textwidth}
         \centering
         \includegraphics[width=\linewidth]{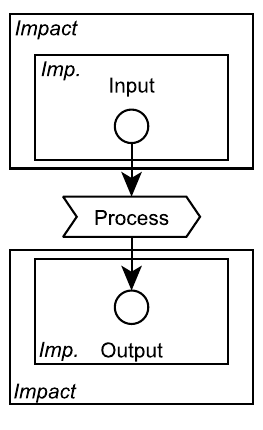}
         \caption{AP~3}
         \label{fig:pattern-3}
     \end{subfigure}
     \hfill
     \begin{subfigure}[b]{0.22\textwidth}
         \centering
         \includegraphics[width=\linewidth]{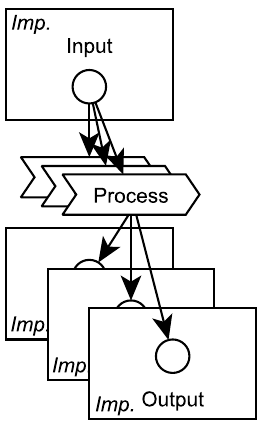}
         \caption{AP~4}
         \label{fig:pattern-4}
     \end{subfigure}
        \caption{Sustainability Anaylsis Patterns}
        \label{fig:patterns}
\end{figure}

\subsubsection*{(AP~1) Sustainability-Relevant Inputs and Outputs:} A process component can be assigned sustainability-relevant inputs and outputs (see Fig.~\ref{fig:pattern-1}). For example, an activity instance requires a certain amount of energy as input or produces a certain amount of waste as output.

%\begin{figure}[htbp]
%\centerline{\includegraphics[width=\linewidth]{pattern-1.pdf}}
%\caption{Assignment of sustainability-relevant inputs and outputs (AP~1).}
%\label{fig:pattern-1}
%\end{figure}

\emph{Motivation:} The recording of inputs and outputs of a business process corresponds to the inventory analysis of LCA. It is a basic requirement to enable reasoning about the sustainability impacts of a process.
% Only through the most complete recording possible of the relevant inputs and outputs of a business process can the significance of its sustainability impacts be determined.

\subsubsection*{(AP~2) Sustainability Impacts:}
A process component can be assigned sustainability impacts (see Fig.~\ref{fig:pattern-2}). Different environmental and social impacts can be distinguished (see the following variants). %The sustainability impacts may be associated with the inputs and outputs (AP 1) assigned to a process component.

%\begin{figure}[htbp]
%\centerline{\includegraphics[width=\linewidth]{pattern-2.pdf}}
%\caption{Assignment of sustainability impacts (AP~2).}
%\label{fig:pattern-2}
%\end{figure}

\emph{AP~2-Env:} A process component can be assigned environmental impacts (e.g., an activity instance %causes a climate impact of 5 kg CO2e or 
contributes to ozone depletion with 3 kg CFCe).

\emph{AP~2-Climate:} A process component can be assigned a climate impact (e.g., an activity instance causes a climate impact of 5 kg CO2e).
%(and specific analysis standards such as the GHG Protocol \cite{wbcsdGreenhouseGasProtocol2004} exist for it).

\emph{AP~2-Social:} A process component can be assigned social impacts (e.g., an activity instance causes work accidents).

\emph{Motivation:} The assignment of impacts to a process component corresponds to the impact assessment in LCA. 
The assessment of impacts is to be differentiated from the mere recording of inputs and outputs.
% The sustainability impacts associated with the inputs and outputs of a process component should be comprehensively determined. 
A comprehensive assessment of impacts allows for identifying shifts between them. AP~2-Climate is defined as a special case of AP~2-Env because climate impacts are particularly relevant in the context of sustainability analysis \cite{wbcsdGreenhouseGasProtocol2004}.

\subsubsection*{(AP~3) Scoping of Sustainability Impacts:}

A process component can be assigned impacts for different scopes (see Fig.~\ref{fig:pattern-3}). This scope definition can be made for different sustainability impacts (see the following variants of AP~3).

%\begin{figure}[htbp]
%\centerline{\includegraphics[width=\linewidth]{pattern-3.pdf}}
%\caption{Scoping of sustainability impacts (AP~3).}
%\label{fig:pattern-3}
%\end{figure}

\emph{AP~3-Env:} A process component can be assigned environmental impacts for different scopes (e.g., an activity instance contributes to ozone depletion with 3 kg CFCe when including direct emissions, and 15 kg CFCe when including emissions in the value chain).
%. The same activity execution contributes to ozone depletion with 15 kg CFCe when including the emissions in the value chain of the used resources.

\emph{AP~3-Climate:} A process component can be assigned climate impacts for different scopes (e.g., following the scoping concept of the GHG Protocol an activity instance causes a Scope 1 impact of 5 kg CO2e and a Scope 3 impact of 30 kg CO2e).
%. The same activity execution causes a Scope 3 climate impact of 30 kg CO2e).
%The definition of the system boundaries may be oriented on Scopes 1-3 of the GHG Protocol \cite{wbcsdGreenhouseGasProtocol2004}. 
% For example, the execution of an activity causes a Scope 1 climate impact of 5 kg CO2e. The same activity execution causes a Scope 3 climate impact of 30 kg CO2e.

\emph{AP~3-Social:} A process component can be assigned social impacts for different scopes (e.g. an activity instance causes an average of 0.00001 work accidents within the company and 0.00002 work accidents when considering the value chain of used resources).

% . For example, executions of an activity cause an average of 0.00001 accidents, including accidents that occur within the considered company. When including accidents that occur during the production of the used resources, the same activity executions cause an average of 0.00002 accidents.

\emph{Motivation:} Following the idea of life cycle thinking, Sustainable BPM should enable the identification of impact shifts along value chains. However, %, depending on data availability and practical concerns, 
fully considering the value chain may not always be feasible. Still, a transparent definition of the considered scopes should be supported for comparability.

\subsubsection*{(AP~4) Allocation of Sustainability Impacts:}
Sustainability impacts assigned to a process component can be (partially) allocated to other process components (see Fig.~\ref{fig:pattern-4}). For example, the impact associated with the production of a resource can be allocated to individual activity instances. % in which the resource is used.

%\begin{figure}[htbp]
%\centerline{\includegraphics[width=\linewidth]{pattern-4.pdf}}
%\caption{Allocation of sustainability impacts (AP~4).}
%\label{fig:pattern-4}
%\end{figure}

\emph{Motivation:} %Challenges in allocating sustainability impacts in multi-input and multi-output processes are addressed in LCA literature. 
To identify improvement potential in a business process, an analysis should yield results on which process components are responsible (to what extent) for various impacts. 
For this, transparent mechanisms are needed to allocate impact assessment between process components.

% For this, the impact assessment may need to be allocated between different process components. For this

%In many cases
% , e.g. when a resource is used by multiple activity instances or an activity instance has multiple outputs,
% A business process may have multiple inputs and outputs. 
% Similar challenges to LCA allocation problems need to be solved in Sustainable BPM, for example, when a process instance has multiple outputs or a resource is used by multiple process instances.

\section{Application of Sustainability Analysis Patterns}
\label{sec:evaluation}

\subsection{Criteria and Overview}

To show the utility of the proposed  sustainability analysis patterns, they are applied to evaluate existing Sustainable BPM approaches.  
The evaluation focuses on modelling approaches since this has been the focus of Sustainable BPM literature \cite{fritschPathwaysGreenerPastures2022}. Corresponding publications were extracted from a literature review on Sustainable BPM \cite{fritschPathwaysGreenerPastures2022}.
In the extracted publications, it can be seen that various authors developed their approaches across several publications. 
% The publications were, therefore, grouped according to the approach described and the authors involved. In the following, each approach is named after the author that appears the most as the primary author in a group of publications. 
% Furthermore, the evaluation is based on the most recent state of the publications on an approach. For example, if earlier publications only describe basic ideas and principles, but concepts and examples for application are formulated later, the approach was evaluated according to the state in the later publications. 
Due to space restrictions, we only reference selected publications for each approach. For an overview of all publications on Sustainable BPM identified in \cite{fritschPathwaysGreenerPastures2022}, see \cite{fritschSustainableBPMPrimary2021}. 
The modelling approaches considered were examined based on 
% the criterion \emph{maturity} and 
the sustainability analysis patterns described in Section~\ref{sec:patterns}. The criteria and their evaluation are explained in the following.

%\subsubsection*{Maturity} The maturity of an approach refers to the level of detail of described concepts and practical applicability. The evaluation levels are based on the maturity levels for technologies proposed in \cite{mankins1995technology}. (\harveyBallFull\ if the approach is implemented with a software tool, \harveyBallHalf\ if concepts and examples for application are formulated, \harveyBallNone\ if basic principles and ideas are described)

%\begin{description}
%    \item[\harveyBallNone] Basic principles and ideas are described. 
%    \item[\harveyBallHalf] Concepts and examples for application are formulated.
%    \item[\harveyBallFull] Feasibility is demonstrated with an implementation in a modelling tool.
%\end{description}

Regarding \emph{AP~1} %To what extent does the approach support assigning sustainability-relevant inputs and outputs to process components? 
an approach is evaluated
\harveyBallFull\, if modelling of inputs/outputs is supported, and \harveyBallNone\ if not.
%\begin{description}
%    \item[\harveyBallNone] No modelling support for AP~1 is provided.
%    \item[\harveyBallFull] Modelling support for AP~1 is provided.
%\end{description}
For \emph{AP~2} %To what extent does the approach support the assignment of sustainability impacts to process components? 
% A distinction is made between whether only climate impacts are considered (AP~2-Climate), environmental impacts including climate impacts (AP~2-Env), or also social impacts (AP~2-Social). 
an approach is evaluated \harveyBallFull\ if modelling of impacts is supported, %\harveyBallHalf\ if impacts can be modelled but are not distinguished from inputs/outputs, 
and \harveyBallNone\ if modelling of impacts is not supported. The evaluation for 
%\begin{description}
%    \item[\harveyBallNone] No modelling support for the assignment of sustainability impacts is provided.
%    \item[\harveyBallHalf] Modelling support for the assignment of sustainability impacts is provided. The approach does not explicitly distinguish between sustainability-relevant inputs/outputs and sustainability impacts.
%    \item[\harveyBallFull] Modelling support for the assignment of sustainability impacts is provided, and inputs/outputs are distinguished from impacts.
%\end{description}
\emph{AP~3} %To what extent does the approach support the scoping of system boundaries when assigning sustainability impacts to process components? 
% A distinction is made between whether only climate impacts are considered (AP~3-Climate), environmental impacts including climate impacts (AP~3-Env), or also social impacts (AP~3-Social). 
distinguishes between 
\harveyBallFull\ if scoping is supported, \harveyBallHalf\ if scoping is considered but explicitly limited, and \harveyBallNone\ if scoping is not supported. 
Finally, for 
%\begin{description}
%    \item[\harveyBallNone] The scoping of impacts is not considered.
%    \item[\harveyBallHalf] The scoping of impacts is considered but explicitly limited.
%%    \item[\harveyBallHalf] Modelling support for the scoping of sustainability impacts is provided. The approach does not explicitly distinguish between sustainability-relevant inputs/outputs and sustainability impacts.
%    \item[\harveyBallFull] Modelling support for the scoping of sustainability impacts is provided.
%\end{description}
\emph{AP~4} %To what extent does the approach support the allocation of sustainability impacts between different process components? (
an approach is evaluated 
\harveyBallFull\ if allocation is supported, and \harveyBallNone\ if not.

%\begin{description}
%    \item[\harveyBallNone] No modelling support for AP~4 is provided.
%    \item[\harveyBallFull] Modelling support for AP~4 is provided.
%\end{description}

\subsection{Evaluation of Sustainable Process Modelling Approaches}

Table~\ref{tab:related_modelling_approaches} provides an overview of existing modelling approaches for Sustainable BPM, the respective modelling languages used, and the evaluation according to the described criteria. 
% The approaches are named after the primary author who appears most frequently in the assigned works.

\begin{table*}[htbp]%
\renewcommand{\arraystretch}{1.2}
\centering
  \caption{Review of existing sustainable process modelling approaches.}
  \label{tab:related_modelling_approaches}
	\begin{tabular}{| l |  c | c c c | c c c | c |}%
  \hline
    \textbf{Approach} &	
    % \textbf{Language} &
    % \textbf{Mat.} &
    \textbf{AP~1} &
    \multicolumn{3}{c |}{\textbf{AP~2}} &
    \multicolumn{3}{c |}{\textbf{AP~3}} &
    \textbf{AP~4}
	\tabularnewline%
    &
    % &
    % &
    &
    Climate &
    Env &
    Social &
    Climate &
    Env &
    Social &
    
    \tabularnewline
    \hline
    Houy et al. & 
    % EPC &
    % \harveyBallNone  & 
    \harveyBallFull & \harveyBallNone & \harveyBallNone & \harveyBallNone & \harveyBallNone & \harveyBallNone & \harveyBallNone & \harveyBallNone
    \tabularnewline
    
    Hoesch-Klohe et al. &  
    % BPMN &
    % \harveyBallHalf  & 
    \harveyBallFull & \harveyBallFull & \harveyBallNone & \harveyBallNone & \harveyBallHalf & \harveyBallNone & \harveyBallNone & \harveyBallFull  
    \tabularnewline
    
    Recker et al. & 
    % BPMN & 
    % \harveyBallHalf  & 
    \harveyBallFull & \harveyBallFull & \harveyBallNone & \harveyBallNone & \harveyBallFull & \harveyBallNone & \harveyBallNone & \harveyBallFull
    \tabularnewline

    Wesumperuma et al. & 
    % BPMN & 
    % \harveyBallHalf  & 
    \harveyBallFull & \harveyBallFull & \harveyBallNone & \harveyBallNone & \harveyBallFull & \harveyBallNone & \harveyBallNone & \harveyBallFull 
    \tabularnewline
    Zhu et al. & 
    % BPMN &
    % \harveyBallHalf  & 
    \harveyBallFull & \harveyBallNone & \harveyBallNone & \harveyBallNone & \harveyBallNone & \harveyBallNone & \harveyBallNone & \harveyBallNone \tabularnewline
    Betz & 
    % XML-nets &
    % \harveyBallNone & 
    \harveyBallNone & \harveyBallFull & \harveyBallFull & \harveyBallFull & \harveyBallFull & \harveyBallFull & \harveyBallFull & \harveyBallNone\tabularnewline
    
    \hline
	\end{tabular}%
\end{table*}

%\begin{table}[htbp]
%\caption{Table Type Styles}
%\begin{center}
%\begin{tabular}{|c|c|c|c|}
%\hline
%\textbf{Table}&\multicolumn{3}{|c|}{\textbf{Table Column Head}} \\
%\cline{2-4} 
%\textbf{Head} & \textbf{\textit{Table column subhead}}& \textbf{\textit{Subhead}}& \textbf{\textit{Subhead}} \\
%\hline
%copy& More table copy$^{\mathrm{a}}$& &  \\
%\hline
%\multicolumn{4}{l}{$^{\mathrm{a}}$Sample of a Table footnote.}
%\end{tabular}
%\label{tab1}
%\end{center}
%\end{table}

%\emph{Houy et al.} \cite{houy_towards_2011,Houy2012a} describe the idea of annotating activities in EPC models with sustainability indicators such as resource consumption or greenhouse gas emissions. A business process could be visually evaluated based on these annotations. 
%%The works describe a first concept, where deeper analysis possibilities are still missing (\harveyBallNone\ for maturity). No analysis capabilities for different process paths or process instances are described (\harveyBallNone\ for analysis). 
%The challenges of allocation and system boundaries are not addressed (\harveyBallNone\ for allocation and system boundaries). Various indicators such as water consumption, energy consumption, and greenhouse gas emissions are mentioned as possibilities for an analysis, but no distinction is made between inventory indicators and impact indicators (\harveyBallNone\ for impact assessment).

\emph{Houy et al.} 
% \cite{houyGreenBPMSustainability2011, Houy2012a} 
\cite{houyGreenBPMSustainability2011} 
describe the idea of annotating activities in EPC models with indicators such as resource consumption or energy demand. 
The publications describe an initial concept. 
% (\harveyBallNone\ for maturity).  
Various sustainability-relevant inputs and outputs, such as water consumption, energy demand, and CO2 emissions, are mentioned as possibilities for annotation (\harveyBallFull\ for AP 1), but the measurement of sustainability impacts is not addressed (\harveyBallNone\ for AP 2). No modelling support is provided for scoping or allocating impacts (\harveyBallNone\ for AP~3 and AP~4).

In the publications of \emph{Hoesch-Klohe et al.} \cite{Hoesch-Klohe2010a}, activities are also annotated with sustainability indicators, in this case, in BPMN process models. 
% Across several publications, 
% \cite{Ghose2010,Hoesch-Klohe2010a,Hoesch-Klohe2011,Hoesch-Klohe2010,Hoesch-Klohe2012,Hoesch-Klohe2012a}, 
% the approach is elaborated from the description of initial ideas and principles to a more detailed description of modelling and analysis possibilities 
%(\harveyBallHalf\ for maturity) 
% \cite{Ghose2010,Hoesch-Klohe2010,Hoesch-Klohe2010a}.  
For the considered indicators, a distinction is made between sustainability-relevant inputs and outputs (such as energy demands) and climate impacts (measured in CO2e) (\harveyBallFull\ for AP~1 and AP~2-Climate). 
% In \cite{Hoesch-Klohe2010}, examples of other environmental impacts and their connection with resource models are mentioned (\harveyBallFull\ for AP~2-Env). 
The consideration of climate impacts is explicitly limited to Scopes 1 and 2 of the GHG Protocol (\harveyBallHalf\ for AP~3-Climate). Other sustainability impacts are not addressed (\harveyBallNone\ for AP~2-Env, AP~2-Social, AP~3-Env and AP~3-Social).
% For other environmental impacts, scoping is not addressed (\harveyBallNone\ for AP~3-Env). Social impacts are also not considered (\harveyBallNone\ for AP~2-Social and AP~3-Social). 
% Regarding allocation, \cite{Ghose2010} suggests annotating individual activities in BPMN business process models with the amount of GHG emissions that occur during the execution of these activities. For this purpose, the resources used in the activities (such as welding equipment or trucks) and their energy demands or fuel consumption should be modelled. With these annotations, the total emissions for different process instances can then be calculated, and design variants with lower consumption can be proposed and examined. 
Regarding allocation, \cite{Hoesch-Klohe2010a} proposes a resource modelling concept (\harveyBallFull\ for AP~4). 
%An example is the resource printer, which is used in an activity and itself requires the resources paper and electricity. 
With these models, the total impacts for different process instances can then be calculated.

In the approach of \emph{Recker et al.} 
% \cite{Recker2011,Recker2012}, 
\cite{Recker2011}, 
a BPMN extension for sustainability aspects is proposed. % including an application example. 
% (\harveyBallHalf\ for maturity). 
With these new notation elements, activities can be marked, as paper-consuming or fuel-consuming (\harveyBallFull\ for AP~1). Furthermore, the resulting climate impact for different groups of activities is displayed in a special notation element (\harveyBallFull\ for AP~2-Climate). 
The calculated climate impacts also consider the production of used resources, i.e. Scope 3 (\harveyBallFull\ for AP~3-Climate). 
% Activity data (for example, the weight of paper or length of distance travelled on a business trip) and emission factors are collected to determine the emission values. %To assess the sustainability impacts of a business process, these determined values are then summed up. 
Concepts for considering additional impacts are not developed (\harveyBallNone\ for AP~2-Env, AP~2-Social, AP~3-Env and AP~3-Social). 
A procedure is proposed in which the climate impacts of resources are allocated (\harveyBallFull\ for AP~4). 
%However, unlike the approach of Hoesch-Klohe et al., this is not model-based. T 

Closely related to \emph{Recker et al.} is the approach of \emph{Wesumperuma et al.} 
% \cite{Wesumperuma2011,Wesumperuma2013}, 
\cite{Wesumperuma2013}, 
which also proposes and elaborates an extension of BPMN with notation elements for capturing climate impacts (\harveyBallFull\ for AP~2-Climate). 
% (\harveyBallHalf\ for maturity). 
A procedure is described in which the climate impact is summed up for individual activities, and processes. Furthermore, a calculation method is proposed that can be used to allocate the sustainability impacts caused by resources to individual activities (\harveyBallFull\ for AP~1 and AP~4). As with \emph{Recker et al.}, the approach considers different scopes of climate impacts, including Scope 3 (\harveyBallFull\ for AP~3-Climate). Other environmental or social impacts are not considered (\harveyBallNone\ for AP~2-Env, AP~2-Social, AP~3-Env and AP~3-Social).

% is limited to determining climate impacts (- for AP 2-Eco and AP 2-Social), but it also considers climate impacts along the value chain (+ for AP 3-Climate, but - for AP 3-Eco and AP 3-Social).

%Closely related to the approach of \emph{Recker et al.} are the works of \emph{Wesumperuma et al.} \cite{Wesumperuma2011,Wesumperuma2013}, where also an extension of BPMN with notation elements for GHG emissions is also proposed. % and worked out with examples (\harveyBallHalf\ for maturity). 
%A procedure is described in which the GHG emissions for individual activities, %sub-processes, 
%and processes are summed up. 
%%However, as with \emph{Recker et al.}, no clear distinction is made between process type and process instance (\harveyBallNone\ for analysis). 
%Furthermore, a calculation procedure is proposed for allocating the sustainability impacts caused by resources to individual activities (\harveyBallHalf\ for allocation). Like \emph{Recker et al.}, the approach is limited to determining GHG emissions (\harveyBallHalf\ for impact assessment), but also considers 'Scope 3' emissions (\harveyBallHalf\ for system boundaries).

The approach of \emph{Zhu et al.} \cite{Zhu2015} involves enriching BPMN process models with sustainability-relevant context data such as resources, countries, or people 
% (\harveyBallHalf\ for maturity, 
(\harveyBallFull\ for AP~1). 
%and then translating them into Colored Petri Nets. % (\harveyBallHalf\ for maturity). 
%The translation into a Colored Petri net allows the model to be simulated (\harveyBallFull\ for analysis). 
In terms of impacts, the approach remains unspecific. It is shown that integration of data from environmental information systems is possible, but the adequate integration of sustainability data in terms of allocation, different impacts, and scoping is not addressed (\harveyBallNone\ for AP~2, AP~3 and AP~4).

%It is proposed to model sustainability aspects in business processes using XML nets. 
%This would also enable the simulation of the model (\harveyBallFull\ for analysis). 
%It is mentioned that various environmental and social indicators can be considered, but there are no concrete concepts provided to address the allocation, indicator or system boundary problem (\harveyBallNone\ for all criteria).

Betz \cite{betzSustainabilityAwareProcess2014} describes the idea of modelling sustainability aspects of business processes with a variant of high-level Petri Nets. It is mentioned that various environmental and social impact indicators can be integrated into a process model. A distinction is made between direct, indirect, and socio-economic impacts (\harveyBallFull\ for AP~2 and AP~3). However, the identification of sustainability-relevant inputs/outputs and allocation are not considered (\harveyBallNone\ for AP 1 and AP 4). 
% Possibilities for allocating sustainability impacts are not described (\harveyBallNone\ for AP 4).
%, as well as further investigations or implementations of the proposal, are not described.
%(\harveyBallNone\ for AP 4 and maturity).

%Several other approaches (\emph{Cappiello et al.} \cite{Pernici2008,Ardagna2008,Cappiello2011,Cappiello2011a,Cappiello2013,Cappiello2014}, \emph{Nowak et al.} \cite{Nowak2013b,Nowak2011a}  and \emph{Reiter et al.} \cite{Reiter2014, Lubbecke2015}) focus on the energy efficiency analysis of IT infrastructure used in processes. Since the developed concepts are limited to IT infrastructure and not transferable to other domains, these approaches are not described in more detail here.

% \subsection{Summary Sustainable BPM }

The identified modelling approaches have only reached a limited degree of maturity. %, in particular, none appears to have been implemented in process modelling tools. 
% The investigation of \cite{stadtlanderHowSoftwarePromotes2019} also concludes that social and environmental analysis support is insufficient in existing business process modelling tools. 
The majority of developed concepts focus on energy aspects and GHG emissions. Apart from measuring climate impacts based on the GHG Protocol, a systematic distinction between inputs/outputs and impacts, as well as concepts for impact scopes, are missing. The approach from \emph{Betz} is an exception with its distinction of direct, indirect and socio-economic impacts. This scoping concept takes an even broader perspective than conventional LCA analyses since it does not only consider the value chain. 
% It can be likened to a special form of LCA analyses, so-called \enquote{consequential} LCA (see, e.g. \cite{ekvallAttributionalConsequentialLCA2016}).
Of the approaches that consider the allocation of sustainability impacts, %(\emph{Hoesch-Klohe et al.}, \emph{Recker et al.}, \emph{Wesumperuma et al.})
\emph{Hoesch-Klohe et al.} appears to be the most sophisticated with detailed resource modelling concepts.

% Regarding modelling support for the allocation of sustainability impacts, three of the approaches (\emph{Hoesch-Klohe et al.) offer elaborated concepts with resource models that enable the distribution of climate impacts. 

\subsection{Evaluation of Sustainable Process Mining Approaches}

So far, sustainability aspects have only been addressed to a limited extent in process mining. 
A systematic literature review on Sustainable Process Mining approaches \cite{gravesReThinkYourProcesses2023} finds that existing publications on the topic provide mainly high-level descriptions. Four case studies identified in the review directly address social or environmental sustainability aspects. 
Three of them, however, only address sustainability-related domains (health and safety \cite{pikaUsingBigData2021}, wind turbine maintenance \cite{duProcessMiningWind2021} and sustainable agriculture \cite{dupuisPredictingCropRotations2022}). The described approaches do not attempt to measure sustainability aspects in processes. Rather, conventional measures (e.g. time usage) are applied. Only \cite{acerbiFosteringCircularManufacturing2022} addresses measuring sustainability aspects in processes by relating energy needs to identified process models (AP~1). 
This shows that further work is required to provide more expressive sustainability analyses with process mining. One solution proposed by \cite{gravesReThinkYourProcesses2023} is to enrich process models with object quantities. This, in turn, would enable future approaches that provide support for allocating impacts between process components (AP~4).

\section{Conclusion and Outlook}
\label{sec:conclusion}

Sustainable BPM approaches need a solid concept of sustainability impact measurement to provide meaningful and comparable analyses. 
The sustainability analysis patterns proposed in this paper provide an initial framework to critically assess Sustainable BPM approaches. 
The mapping provided in Section~\ref{sec:mapping} shows that Sustainable BPM and LCA share common goals and concepts. Therefore, in the future,  further insights from LCA can be adapted for Sustainable BPM to elaborate and extend the proposed patterns.
% On the one hand, future Sustainable BPM research should provide further guidelines and conventions. On the other hand, there is room for exploring different %and varying 
% impact concepts in Sustainable BPM research. 
The LCA standard \cite{isoISO1404020062006} primarily defines certain principles but does not prescribe specific techniques. In this sense, Sustainable BPM approaches can be seen as a technical variant of other LCA approaches.
% BPM doesn't need to reinvent the wheel regarding impact assessment.
% Will BPM develop its own notion of impacts? Maybe a set of conventions may make sense (only cradle-to-gate?)
% There is room vor variety (as is in LCA) - maybe LCA merges with BPM, maybe, in the future BPM may develop an own concept. 
% Special Perspective/Contribution of Sustainable BPM
% Different impact assessment concepts may emerge, similar to LCA attributional versus consequential.
% In this modelling perspective, process models do not represent different execution paths. The process models are less built for (discrete) process simulation but rather serve as a basis for building equation systems. 
% Advantages of each of the modelling perspectives 
However, the BPM (modelling) perspective provides specific advantages regarding the continuous support of process changes, as well as the implementation of tools and IT support \cite{weskeBusinessProcessManagement2019}. 
% We, therefore, believe that 
Sustainable BPM thus brings a unique and promising contribution to effectively improving the sustainability performance of companies. 

% The BPM perspective on processes is not only useful for analysing 

% , specifically since BPM it is not only concerned with analysis, but also may provide tools for implementing changes, and specifically IT support, automatisation, digitalisation. 

%\section*{Acknowledgment}
%
%The preferred spelling of the word ``acknowledgment'' in America is without 
%an ``e'' after the ``g''. Avoid the stilted expression ``one of us (R. B. 
%G.) thanks $\ldots$''. Instead, try ``R. B. G. thanks$\ldots$''. Put sponsor 
%acknowledgments in the unnumbered footnote on the first page.

\bibliographystyle{splncs04}
\bibliography{main}

\end{document}